# Global Cold Curve.
# New representation for Zero-temperature isotherm in whole density range

Iosilevskiy I.L.

*Moscow Institute of Physics and Technology (State University), Dolgoprudny, Russia*

ilios@orc.ru

Improved representation for so-called "cold curve" (CC) /i.e. isotherm $T = 0$/ for gaseous plasma state is necessary for a long time. In standard (traditional) representation for CC the dependence of internal energy $U$ on mass-density $\rho$ (or specific volume $v$) is used at $T = 0$ in whole density range, $U = U(\rho)$ The main drawback of such representation for CC [1, 2] is total absence in this form of physically meaningful part for gaseous state and presence of physically meaningless part instead, which corresponds to absolutely unstable (labile) thermodynamic states in whole gas-like low density range, i.e. at $\rho < \rho_{Sp}$ (here $\rho_{Sp}$ is the spinodal density of condensed state). It is illustrated at Fig.1 (see also [3] and Fig.20 at p.370 in [2])

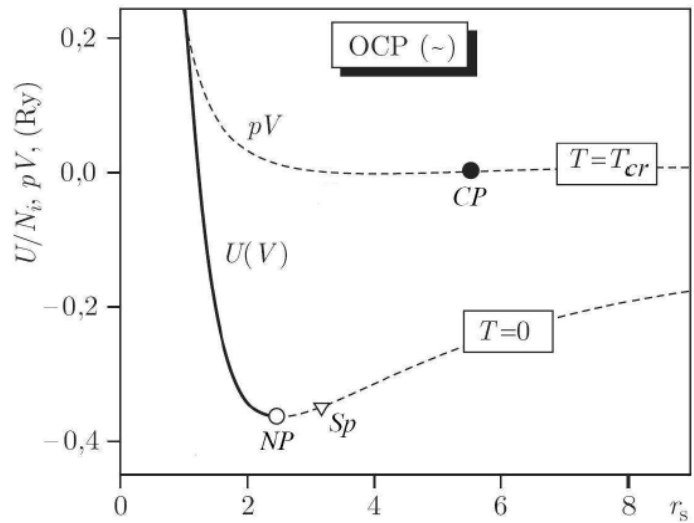

**Figure 1**. Standard cold curve ($T = 0$) and critical isotherm ($T = T_{cr}$) in modified one-component plasma model OCP(~) [3, 2]. Notations: $U$, $V$, $p$ – internal energy, volume and pressure; $r_s$ – specific volume in atomic units ($r_s \equiv a/a_B$ / n$^{-1/3}$); NP – "normal point" ($T = 0$, $p = 0$); CP – critical point; $Sp$ – spinodal ($\partial p/\partial V = 0$); $0 < r_s < NP$ – stable part of CC; $NP < r_s < Sp$ – metastable part of CC (expanded crystal); $Sp < r_s < \infty$ – unstable (meaningless) part of standard CC.

It is very important because traditional form of cold curve plays significant role in all wide-range approximations for EOS of gaseous and plasma state, including critical point of gas-liquid phase transition. In these approximations final EOS is constructed as algebraic envelope (with free parameters), which is the sum of standard cold curve $U(\rho)$ with additional thermal part $\Delta U(\rho,T)$. It should be stressed that location and properties of the critical point in these semi-empirical EOS could be very sensitive to properties of mentioned above meaningless non-stable part of traditional cold curve.

Non-standard representation for generalized cold curve (GCC) was proposed in [4, 3]. This proposal based on the study of remarkable structure of thermodynamic properties of gas-like plasma in joint limit $T \to 0 \cap \rho \to 0$ [5] (see also [2, 6]).

The main point of new representation is that chemical potential of substance, μ, but not density, plays role of ruling parameter in proposed new GCC dependence $U = U(\mu)$, in contrast to the standard form $U = U(\rho)$. This substitution changes radically low-density (gaseous) part of GCC [2, 3]. Namely:

(1) Physically meaningless part of standard cold curve $U(\rho)$ *disappears* from new GCC, $U = U(\mu)$.

(2) New totally meaningful branch of gaseous cold curve comes in GCC. It corresponds to thermodynamically stable states only.

(3) The new gaseous branch of GCC describes in simple, schematic way *all* thermodynamic events in gas-like plasma in joint limit: $T \to 0 \cap \rho \to 0$, in contrast to standard representation $U = U(\rho)$, which describes atomic state only. I.e. the new branch is combination of *all ionization* and *dissociation* processes available for rarefied equilibrium gaseous
plasma at finite temperature [2 - 5].

(4) The new gaseous branch of GCC located at μ-axis in totally negative interval. Thermodynamically stable part of GCC is limited below by the major ionization potential value $-Z^2 Ry$, and above – by the value of chemical potential in saturation point of gas-crystal transition (sublimation) [1, 2]:

$$(\mu_{el})_{sub} > \mu_{el} > (\mu_{el})_{min} = -Z^2 Ry. \tag{1}$$

(5) Another meaningful portion of the zero-temperature isotherm corresponds to metastable gaseous state (supercooled vapor). It is also located at negative side of chemical potential axis. It is limited above by the (unknown) chemical potential value in *spinodal* point of gaseous state:

$$(\mu_{el})_{spin} > \mu_{el} > (\mu_{el})_{sub} \tag{2}$$

(6) The binding energies of *all* available complexes (atomic, molecular, ionic and even clusters (!) in their *ground state* with addition of the sublimation energy of the gas-crystal phase transition are the only quantities that display themselves in meaningful details of this new generalized gaseous part for zero-temperature isotherm (so-called "energy scale" at GCC [1 - 4]).

(7) New generalized representation of zero-temperature thermodynamics is possible (and meaningful) not only for caloric EOS (internal energy vs. chemical potential $U = U(\mu)$ but for *thermal* EOS also in form of dependence for compressibility factor ($PV/RT$) vs. chemical potential [1, 2] (see Fig. 2)

**Thermodynamics of rarefied gaseous plasma in zero-temperature limit.**

A knowledge of plasma thermodynamics proves to be asymptotically exact within three well-known limits, the ideal nucleus-electron mixture being the reference:

A: $n = const \cap T \to inf$ – Thermal ionization (A)
B: $T = const \cap n \to inf$ – Pressure ionization (B)
C: $T = const \cap n \to 0$ – Expansion ionization (C)

There is one more limit where our knowledge of plasma thermodynamics also looks asymptotically exact:

D: $T \to 0 \cap n \to 0$ (D)

Thermodynamic functions of gaseous plasmas in the limit of extremely low temperature and density obtain remarkable simple and schematic structure. The tendency, which was claimed previously [7], is carried to extreme.

It should be stressed that the point $T = 0$ and $n = 0$ is singular one. In this case matter thermodynamics depends on the kind of limiting process, i.e.:

$$\lim_{T\to 0} \lim_{n\to 0} \neq \lim_{n\to 0} \lim_{T\to 0} \tag{L}$$

The point is that discussed limit ($T \to 0 \cap n \to 0$) is carried out at fixed value for chemical potential of electrons ($\mu_{el} = const$) or atoms ($\mu_a = const$) or molecule ($\mu_{mol} = const$) etc.

In this limit both equations of state (EOS), thermal and caloric ones, obtain almost identical stepped structure ("ionization stairs" [4, 5]) when one uses special forms for exposition of these EOS as a function of electron chemical potential: i.e. $PV/RT$ for thermal EOS and $U - (3/2)PV$ for caloric EOS vs. $\mu_{el}$. Examples of this limiting structure are exposed at figures 1 and 2 in [6] for thermal and caloric EOS of lithium and helium plasmas (see also [1, 2]). For rigorous theoretical proof of existing the limit, which is under discussion (SAHA-limit) in the case of hydrogen, see [12, 13] and references therein.

The same stepped structure is valid in the zero-temperature limit for any molecular gases, for example for hydrogen (see Fig. 3 in [6]). It is also valid for simple plasma mixtures, for example for zero-temperature hydrogen-helium mixture with typical Sun-like composition (see below). This limiting structure appears within a fixed (negative) range of ($\mu_{el}^{**} \geq \mu_{el} \geq \mu_{el}^{*}$). It is bounded below by value of major ionization potential of given chemical element ($\mu_{el}^{*} = -I_Z = -Z^2 Ry$) and above by the value depending on ionization potential and sublimation energy of substance $\{\mu_{el}^{**} = (\Delta_s H^o + I_1)/2\}$ [1,2]. Binding energies of all possible bound complexes (atomic, molecular, ionic and clustered) in its *ground state* are the *only* quantities that manifest itself in meaningful details of this limiting picture as location and value of every step. The energy of *macroscopic binding* – the heat of condensation at $T = 0$ – supplement this collection. At the same time there are *no such steps* for *exited states* of all bounded complexes (ions, atoms, molecules and clusters). Altogether, all energies mentioned above form *intrinsic energy scale* [1, 2]. It could be considered as energy "passport" for any substance.

In the zero-temperature limit all thermodynamic differential parameters (heat capacity, compressibility, etc.) obtain their remarkable δ-like structures ("*thermodynamic spectrum*" [1, 2]). Both kinds of such "spectrum" became apparent: i.e. "emission-like spectrum" for heat capacity (see Fig.5 in [6]) and "absorption-like spectrum" for the isentropic coefficient $\gamma_S \equiv (\partial \ln p/\partial \ln \rho)_S$ (see Fig.6 in [6]). It should be stressed that all "*lines*" of these "thermodynamic spectrum" are *centralized* just at the elements of the "intrinsic energy scale" – binding energies of ground states for all bound complexes in the system.

The limiting EOS stepped structure ("ionization stairs") of gaseous zero-$T$ isotherm is generic prototype [7] of well-known "shell oscillations" in EOS of gaseous plasmas at low, but finite temperatures and non-idealities [1, 2]. At the same time this limiting stepped-like zero-temperature form of gaseous plasma thermodynamics for Coulomb systems could be used as a natural basis for rigorous deduction of well-known quasi-chemical approach ("chemical picture") in frames of asymptotic expansion around this reference system. The point is that this expansion must be provided on *temperature* at *fixed chemical potential* (see below), in contrast to the standard procedure of expansion on *density* at *constant temperature* [14, 15].

**Generalized zero-temperature isotherm for gaseous mixtures.**

Remarkable step-like structure of zero-$T$ isotherm for single-element cold curve like exposed at Fig.1, could be generalized naturally for cold "plane" of two- or more element mixture at $T = 0$. This generalized cold plane is still cold curve when we use electron chemical potential $\mu_{el}$ as the ruling parameter. As an example cold curve for the Sun-like He + $H_2$ mixture is exposed at Figure 2.

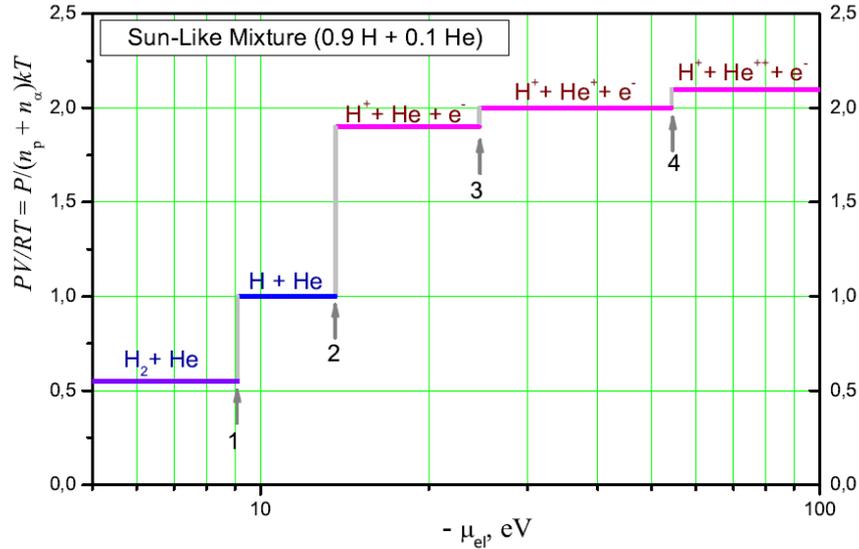

**Figure 2**. Generalized cold curve for the Sun-like mixture (0.9 H + 0.1He). Compressibility factor (*PV/RT*) vs. electronic chemical potential at *T* = 0. Notations: *1, 2* – dissociation and ionization of hydrogen; *3, 4* – first and second ionizations of He.

**Complex gas-crystal cold curve.**

Gaseous portion of new generalized zero-temperature isotherm can be naturally conjugated with corresponding cold curve of condensed state [3]. This new combined two-part zero-temperature isotherm does not contain any more the meaningless part, corresponding to thermodynamically unstable states (i.e. artificial portion of cold curve between gaseous and crystalline spinodals). Example of such complex gas-crystal cold curve is exposed at Figure 3.

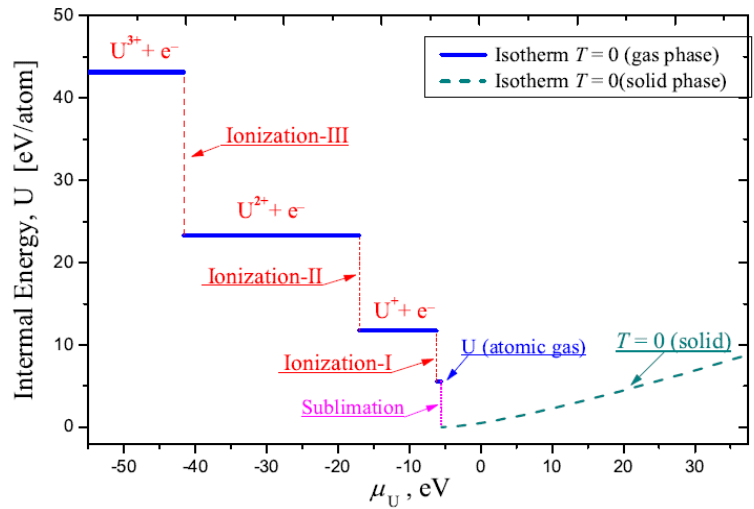

**Figure 3**. Generalized gas-crystal cold curve for Uranium. Gaseous (left) and crystal (right) portions of complex cold curve. Modified form $U = U(\mu)$ for crystal was recalculated presently from traditional form $U = U(\rho)$, which was kindly given to the author by I.V. Lomonosov [9])

**Step-like structure of cold curve for matter in ultra-high density limit.**

The simple, schematic structure ("ionization stairs"), which is typical for new form of cold curve in ultra-low densities, appears again at zero-temperature isotherm in ultra-high densities of matter in conditions, which are typical for interiors of compact stars (neutron star crust) [8]. Adequate theoretical approaches [8] consider this state of matter as strongly interacting zero-temperature mixture of protons, neutrons, electrons and nuclei {*p, n, e* and *N(A,Z)*} under equilibrium reactions of nuclei creation and decomposition (3) and *β*-equilibrium (4):

$$N(A,Z) = An + Zp \tag{3}$$
$$p + e = n + v \tag{4}$$

The discussed step-like structure of matter cold curve in ultra-high densities is illustrated at Figure 4 from [8]. We can conclude that in new generalized representation the low-temperature thermodynamics of matter in ultra-low and ultra-high densities have many features of the same remarkable structure by the same physical reason.

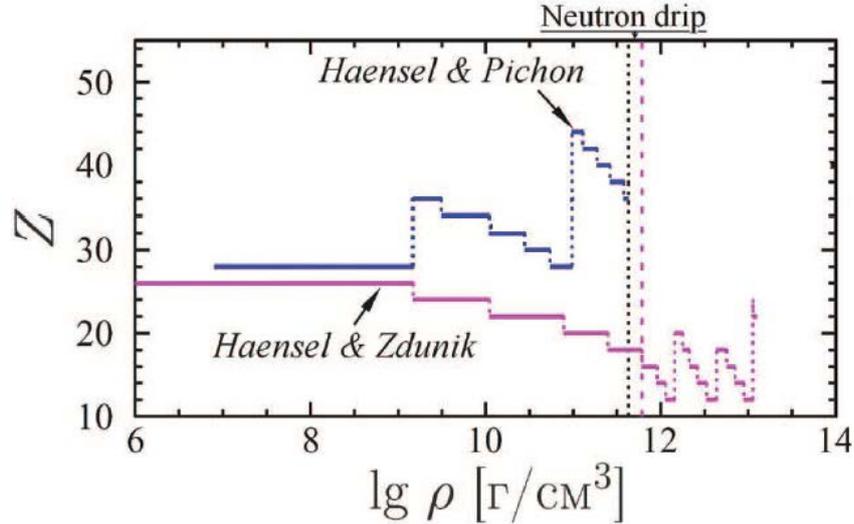

**Figure 4**. Mono-nuclear layers in neutron star crust [8] in equilibrium zero-temperature conditions [10] and in conditions of accreting neutron star crust [11].

**Acknowledgments**.
  The work was supported by Grant ISTC 3755, by RAS Scientific Program "Research of matter under extreme conditions" and by MIPT Education Center "Physics of High Energy Density Matter".